\documentclass[prd,superscriptaddress,onecolumn,showkeys]{revtex4}
\usepackage{textcomp}
\usepackage{eurosym}
\usepackage{amsfonts}
\usepackage{array}
\usepackage{amsthm}
\usepackage{bm}
\usepackage{palatino}
\usepackage[colorinlistoftodos]{todonotes}
\usepackage{mathpazo}
\usepackage{supertabular}
\usepackage{subfig}
\usepackage{amssymb}
\usepackage{eurosym}
\usepackage{amsmath}
\usepackage{epsfig}
\usepackage{float}
\usepackage{graphics}
\usepackage{changes}
\usepackage{color}
\usepackage{graphicx}
\usepackage[colorlinks=true,
            linkcolor=blue,
           urlcolor=black,
           citecolor=black]{hyperref}
\def\be{\begin{equation}}
\def\ee{\end{equation}}
\def\beq{\begin{eqnarray}}
\def\eeq{\end{eqnarray}}

\def\bes{\begin{eqnarray}}
\def\ees{\end{eqnarray}}

\begin{document}
\title{First-order correction of tunneling and entropy in the Horndeski gravity-like hairy black hole}

\author{Riasat Ali}
\email{riasatyasin@gmail.com}
\affiliation{Department of Mathematics, Shanghai University and Newtouch Center for Mathematics of Shanghai University, Shanghai-200444, People's Republic of China}

\author{Xia Tiecheng}
\email{xiatc@shu.edu.cn}
\affiliation{Department of Mathematics, Shanghai University and Newtouch Center for Mathematics of Shanghai University, Shanghai-200444, People's Republic of China}

\author{Rimsha Babar}
\email{rimsha.babar10@gmail.com}
\affiliation{Department of Mathematics, GC
University Faisalabad Layyah Campus, Layyah-31200, Pakistan}

\date{\today}

\begin{abstract}
In this work, we apply tunneling formalism to analyze charged particles tunneling across a hairy black hole horizon. Such black hole solutions are essential for frameworks based on Horndeski's gravity theory. Applying a semi-classical technique, we examine the tunneling of charged particles from a hairy black hole and derive the generic tunneling spectrum of released particles, ignoring self-gravitational and interaction. It is studied to ignore the back-reaction impact of the radiated particle on the hairy black hole. We analyze the properties of the black hole, such as temperature and entropy, under the influence of quantum gravity and also observe that the first-order correction is present. We study tunneling radiation produced by a charged field equation in the presence of a generalized uncertainty effect. We modify the semi-classical technique by using the generalized uncertainty principle, the WKB approximation, and surface gravity.
\end{abstract}

\keywords{Horndeski like hairy black hole; Hawking temperature; semi-classical approach; Quantum correction; corrected entropy}

\date{\today}

\maketitle

\section{Introduction}

A black hole (BH) is an object that absorbs all matter and energy from surrounding space due to its immense gravitational field. According to general relativity (GR), BH absorbs any particle that clashes with a BH horizon.  In $1974$, Hawking give the idea \cite{A1, A2} that a BH could emit thermal emission; multiple studies have been conducted to study different aspects of this phenomenon known as Hawking radiation. Black holes are predicted to be able to lose energy as thermal emissions shrink and ultimately evaporate by this analysis, which is a quantum effect. Several efforts have been made to examine this emission spectrum from BHs by considering the quantum theories of Dirac, fermion, and light particles, among others. To learn more about the emission spectrum and temperature of various BHs, several researchers \cite{A3, A4, A5, A6, A7, A8, A9, A10} have investigated particle tunneling. The particle tunneling from Kerr-Newman BH has realized significant advances in BH physics \cite{A11} and the black or cosmic string \cite{A12, A12a}. 

The study of fermions tunneling from Reissner-Nordstrom-de Sitter BH with a monopole parameter \cite{A13} employs the WKB approximation and field equation to assess particle temperature and tunneling mechanism. The authors of this study have calculated the temperature and the tunneling probability for fermion tunneling from the horizon. The temperature of a departing fermion from the event horizon and its graphical behavior determines the tunneling procedure for Plebanski-Demianski BHs \cite{A14}. The temperature of Newman-Unti-Tamburino-like BH solutions in the field equations was examined using acceleration and rotation. Quantum theory on a curved background allows for the small-scale measurement of a BH \cite{A15}. For an outgoing particle, the imaginary part of the particle's action examines the tunneling probability. Calculating the tunneling of charged and uncharged scalar and Dirac particles with various BH configurations has been attempted numerous times \cite{A16, A17, A18, A19, A20, A21, A22, A23, A24}. The Unruh temperature has been computed, and the tunneling of particles by the Rindler spaca Kalb-Ramond-like BH's thermodynamics, including Hawking temperature and entropy, wking radiation as a tunneling event using a semi-classical method. The computation for the process of $s$-wave production across the event horizon is part of this procedure. It has been demonstrated in \cite{A28} that a rotating wormhole's Hawking radiation may release all kinds of particles.
Further study has been done on the physical analysis and physics of novel Schwarzschild BHs \cite{A29, A30, A31}. According to the impact of relativistic gravity, the tunneling method has been employed to address the Hawking temperatures for various BH types \cite{A32, A33, A34}. The physical properties of BHs have been successfully developed since the discovery of Hawking radiation using quantum theory in curved spacetime.

The Kalb-Ramond solutions used the Newman-Janis technique, which includes the spin parameter, and also computed the temperature of a Kalb-Ramond BH. A quantum tunneling technique based on surface gravity has been adopted. It has also investigated the thermodynamics, including Hawking temperature and entropy, of a Kalb-Ramond-like BH using first-order quantum correction. A semi-classical approach with the Lagrangian equation in the WKB approximation has been studied for the adjusted temperature in \cite{A35}. The link between BH entropy and horizon area was found in \cite{A36}. Several researchers have suggested various ways to examine standard temperature \cite{A36a, A36b, A36c, A36d} in different BH geometry. The Hawking spectrum of fermions released by a black hole that is minimally geometrically distorted has been examined in \cite{A36f}. Quantum impacts on the geometry are naturally described by the minimally geometrically distorted length scale connected, for example, to extra dimensions. It is concluded that, depending on the relative strength of the minimally geometrically distorted and GUP parameters, the total flux of fermions released by the black hole can disappear for a critical mass.

The tunneling radiation generated by the Hawking idea of charged boson particles from a BH horizon is the subject of this research. Particles have radiated from the hairy BH outer horizon by a semi-classical phenomenon known as tunneling. This analysis focuses on the imaginary part of classical action, which causes boson particles to appear from Horndeski-like hairy BH by tunneling. A thorough examination reveals the intense curiosity of various BHs about the Hawking temperature using the tunneling approach. In the Standard Model, spin-$1$ charged bosons play an essential part. The Lagrangian expression can be used to determine the behavior of the bosons in the background of BH geometry and quantum gravity. First, we use the Glashow-Weinberg-Salam model's Lagrangian to formulate the field equations under the influence of charged bosons' general uncertainty principle (GUP). Next, we will examine the particle emission process by applying the WKB approximation and Hamilton-Jacobi process to the resulting expression for the charged case in the examined Horndeski gravity-like hairy BH. The temperature of a Horndeski gravity-like hairy BH meets the GUP criteria for maintaining its physical and stable state. To fulfill the GUP condition, the next-order corrections must be lower than the standard term of uncertainty expression. The results indicate a positive temperature value that coincides with the GUP expression. Without these conditions, the temperature becomes very negative or very high, indicating the non-physical state of Horndeski's gravity-like hairy BH. The effect of the GUP on Hawking temperature was important \cite{A42} because it implies that the first-order temperature of radiation emitted by Horndeski's gravity-like hairy BH radiation can be changed when taking quantum gravity effects into account. This could result in a lower temperature than the standard calculation and provide insights into the nature of Horndeski's gravity-like hairy BH at the quantum level and the kinetic energy of particles and quantum gravity introduced into Hawking temperature, changing how particles can escape a BH and affecting its temperature. In the context of GUP, we begin with the kinetic term of the charged field in flat spacetime. We extend this to the case of charged BH spacetime with a charged vector boson field. Although other transformations are complicated, the additional derivatives also affect the local unitary (linear) transformation operator $U(x)$ \cite{A42} when taking the gauge principle into account. The WKB approximation states that the fundamental universal constant that defines the quantum nature of energy and connects a photon's energy to its frequency is $6.62607015\times 10^{-34}$ joule-hertz$^{-1}$. This constant constitutes only the smallest order in Planck's constant, which corresponds to the symbol $h$. In \cite{A36g, A41}, we disregard all estimates of the second or higher order Planck's constant because it is very small effect to preserve it in the WKB approximation.

The format of the paper is as outlined below: we start Sec. {\bf II} by reviewing thoughts associated with the Horndeski gravity, like hairy BH solution and Hawking temperature ($T_H$). In Sec. {\bf III}, we examine the improved Hawking temperature ($T'_H$) for Horndeski gravity-like hairy BH in the presence of GUP and graphical analysis of the $T'_H$. In Sec. {\bf IV}, we examine the improved entropy of the BH. At last, we summarise our conclusion in {\bf V}.

\section{INTRODUCTORY REVIEW OF HORNDESKI GRAVITY LIKE HAIRy BLACK HOLE}

Scalar fields in Horndeski's theory can alter spacetime geometry, so solutions to field equations differ significantly from those in GR. The hairy BH, in particular, is a solution in which the BH is surrounded by scalar hairs or extra degrees of freedom (scalar fields) that influence the black hole's characteristics. These "hairy" solutions are crucial because, in normal general relativity, their mass and charge uniquely define BH and angular momentum (the no-hair theorem). Adding scalar fields results in BH solutions with non-trivial features, such as scalar hair, which may exhibit features not found in classical BHs, such as changed event horizons or modified spacetime surrounding the BH. Understanding these solutions can provide fresh insights about gravity, particularly in the context of modified theories. The Horndeski theory's hairy BHs could shed light on the characteristics of dark matter, dark energy, and the early cosmos. These answers could change our knowledge of cosmic development by assisting in investigating the large-scale interactions between other fields and gravitational fields. This portion provides an overview of a metric for a Horndeski gravity-like hairy BH. Horndeski gravity is a four-dimensional scalar-tensor framework that leads to second-order field expression, a version of GR \cite{A37, A38}. The action is explained as follows:
\begin{equation}
I=\int \textrm{d}^{4}y\sqrt{-g}\sum_{i=2}^{5}{\L}_{i},\label{A1}
\end{equation}
with $\L_{i}$ Lagrangian defined as
\begin{eqnarray}
Y&=&-\frac{1}{2}\nabla_{\mu}\psi\nabla^{\mu}\psi,\\
\L_{2}&=&F_{2}(\psi ,Y),\nonumber\\
\L_{3}&=&-F_{3}(\psi ,Y)\Box\psi,\\
\L_{4}&=&F_{4}(\psi ,Y)R+F_{4, Y}(\psi ,Y)\left[(\Box\psi)^{2}-\nabla_{\mu}\nabla_{\nu}\psi \nabla^{\mu}\nabla^{\nu}\psi\right],\\
\L_{5}&=&F_{5}(\psi ,Y)F_{\mu\nu}\nabla^{\mu}\nabla^{\nu}\psi-\frac{1}{6}F_{5, Y}(\psi, Y)\nonumber\\&\times&\left[(\Box \psi)^{3}+2\nabla_{\nu}\nabla_{\mu}\psi\nabla^{\nu}\nabla^{\lambda}\psi\nabla_{\lambda}\nabla^{\mu}\psi-3\Box\psi\nabla_{\mu}\nabla_{\nu}\psi\nabla^{\mu}\nabla^{\nu}\psi\right],
\end{eqnarray}
the scalar field $\psi$ and its kinetic component are arbitrary functions of $F_{i}(\psi, Y)$.
We examine a subclass of the theory after \cite{A39} that views $F_{i}$ as simply a function of the kinetic component, i.e., $F_{i}(Y)$ and $F_{5}(Y)=0$. The action (\ref{A1}) can be varied about the metric to yield the field equations
\begin{equation}
F_{4}(Y)F_{\mu\nu}=T_{\mu\nu},\label{A2}
\end{equation}
with
\begin{eqnarray}
T_{\mu\nu}&=&\frac{1}{2}\left(F_{2,Y}\nabla_{\mu}\psi\nabla_{\nu}\psi+F_{2}g_{\mu\nu}\right)
+\frac{1}{2}F_{3, Y}\left(\nabla_{\mu}\psi\nabla_{\nu}\psi\Box\psi-g_{\mu\nu}\nabla_{\alpha}Y\nabla^{\alpha}\psi-2\nabla_{(\mu}Y\nabla_{\nu)}\psi\right)\nonumber\\
&-& F_{4,Y}\Big[\nabla_{\gamma}\nabla_{\mu}\psi\nabla^{\gamma}\nabla_{\nu}\psi-\nabla_{\mu}\nabla_{\nu}\psi\Box\psi+\frac{1}{2}g_{\mu\nu}\left((\Box\psi)^{2}-(\nabla_{\alpha}\nabla_{\beta}\psi)^{2}-2R_{\sigma\gamma}\nabla^{\sigma}\psi\nabla^{\gamma}\psi\right)\nonumber\\
&-&\frac{R}{2}\nabla_{\mu}\psi\nabla_{\nu}\psi+2R_{\sigma}(\mu|\nabla^{\sigma}\psi\nabla_{|\nu)}\psi+R_{\sigma\nu\gamma\mu}\nabla^{\sigma}\psi\nabla^{\gamma}\psi\Big]- F_{4,YY}\Big[g_{\mu\nu}\left(\nabla_{\alpha}Y\nabla^{\alpha}\psi\Box\psi+\nabla_{\alpha}Y\nabla^{\alpha}Y\right)\nonumber\\
&+&\frac{1}{2}\nabla_{\mu}\psi\nabla_{\nu}\psi\left((\nabla_{\alpha}\nabla_{\beta}\psi)^{2}-(\Box\psi)^{2}\right)-\nabla_{\mu}Y\nabla_{\nu}Y-2\Box\psi\nabla_{(\mu}Y\nabla_{\nu)}\psi-\nabla_{\gamma}Y\nonumber\\&\times&\left(\nabla^{\gamma}\psi\nabla_{\mu}\nabla_{\nu}\psi-2\nabla^{\gamma}\nabla_{(\mu}\psi\nabla_{\nu)}\psi\right)\Big]. \label{A3}
\end{eqnarray}
The definition of the finite four-current $K^{\mu}$, which indicates the scalar field's invariance given shift symmetry as
\begin{equation}
K^{\nu}=\frac{\delta S}{(\sqrt{-g})\delta\psi_{,\mu}},\label{A4}
\end{equation}
and in this instance, it says
\begin{eqnarray}
K^{\nu}&=&-F_{2, Y}\psi^{,\nu}-F_{3, Y}\left(\psi^{,\nu}\Box\psi+Y^{,\nu}\right)-F_{4,Y}\left(\psi^{,\nu}R-2R^{\nu\sigma}\psi_{,\sigma}\right)
-F_{4, YY}\Big[\psi^{,\nu}\left((\Box\psi)^{2}-\nabla_{\alpha}\nabla_{\beta}\psi\nabla^{\alpha}\nabla^{\beta}\psi\right)\nonumber\\
&+&2\left(Y^{,\nu}\Box Y-Y_{,\mu}\nabla^{\mu}\nabla^{\nu}\psi\right)\Big].\label{A5}
\end{eqnarray}
For the metric
\begin{equation}
\textrm{d}s^{2}=-g_{tt}\textrm{d}t^{2}+\frac{\textrm{d}r^{2}}{g_{rr}}+r^{2}\textrm{d}\theta^{2}+r^{2}\sin^{2}\theta \textrm{d}\psi^{2},\label{A6}
\end{equation}
the previously described current's non-vanishing element has a specific form as
\begin{equation}
K^{r}=-F_{2, Y}g_{rr}\psi^{\prime}-F_{3, Y}\frac{4g_{tt}+rg^{\prime}_{tt}}{2rg_{tt}}+2F_{4,Y}\frac{g_{rr}}{r^{2}g_{tt}}\left[(g_{rr}-1)g_{tt}+rg_{rr}g^{\prime}_{tt}\right]\psi^{\prime}
-2F_{4, YY}\frac{g^{3}_{rr}(g_{tt}+rg^{\prime}_{tt})}{r^{2}g_{tt}}\psi^{\prime 3},\label{A7}
\end{equation}
where $(\prime)$ stands for the radial coordinate differentiation. To simplify, we set
\begin{eqnarray}
F_{2}&=&\alpha_{21}Y+\alpha_{22}(-Y)^{\omega_2},\\
F_{3}&=&\alpha_{31}(-Y)^{\omega_3},\\
F_{4}&=&\frac{1}{8\pi}+\alpha_{42}(-Y)^{\omega_4}.
\end{eqnarray}
Given the existence of hairy approaches, we set
\begin{equation}
\omega_2=\frac{3}{2},~~~~~\omega_4=\frac{1}{2},~~~~~\alpha_{21}=\alpha_{31}=0,
\end{equation}
and for considering $K^{\prime}=0$, we get
\begin{equation}
\psi^{\prime}=\pm \frac{2}{r}\sqrt{\frac{-\alpha_{42}}{3g_{rr}\alpha_{22}}}.
\end{equation}
Check \cite{A39} to view the full derivation, and the metric provides us with
\begin{equation}
g_{tt}=g_{rr}=1-\frac{2M}{r}+\frac{Q}{r}\ln\frac{r}{2M},
\end{equation}
here, $Q$ and $M$ indicate the charge and mass connected to the non-trivial field through the length dimension.
We calculate every higher-order term and complete the study of spin-$1$ bosons. This, along with earlier findings \cite{A39a} for fermions and scalars, provides the definitive evidence that Hawking radiation is back-reaction-independent. Although our findings demonstrate that bosons are released at the Hawking temperature, independent of the basic concept's symmetries $T_H=\frac{g_{tt}^{\prime}(r_{+})}{4\pi}$, the computation is without the back-reaction. We get the following Hawking temperature:

\begin{equation}\label{TH1}
T_{H}=\frac{M}{2\pi r_{+}^{2}}+\frac{Q}{4\pi r_{+}^{2}}-\frac{Q}{4\pi r_{+}^{2}}\ln\frac{r_{+}}{2M},
\end{equation}
for each kind of particle. This corresponds with the temperature often given in the literature and is typically only computed according to the leading order of $h$ (Planks constant).
The computation of the temperature from a standard has been a contentious topic for a long time because of the numerous concerns about the strategy's variance. Specifically, it seems to have a distinct Hawking temperature effect by a factor of three: mass, charge, and horizon radius, as well as a logarithmic function based upon the coordinate system ($t,~r,~ \theta,~\psi$) employed. Furthermore, the Schwarzschild BH of temperature 
\Big(i.e., $T_{H}=\frac{1}{8\pi M}$\Big) is recovered when $r_{+}=2M$ and the charge is disregarded. Most people believe we have accurate information on this problem because the tunneling approach has developed and become more understood in BH spacetime. Scalar fields may modify the geometry of a BH, affecting how heat is released from it. The alteration of spacetime outside the event horizon (due to the interaction of the scalar field with the gravitational field) may result in changes to the emission spectrum, including a different temperature for the BH. The presence of scalar hair (the scalar field) around the BH can alter the Hawking radiation compared to that released by conventional BHs. The scalar field can change the temperature of a BH by influencing its event horizon structure. Specifically, the temperature may diverge from the standard shape anticipated by Hawking's theory of BHs in GR.

\begin{figure}[H]
\centering
\includegraphics[width=6cm,height=6cm]{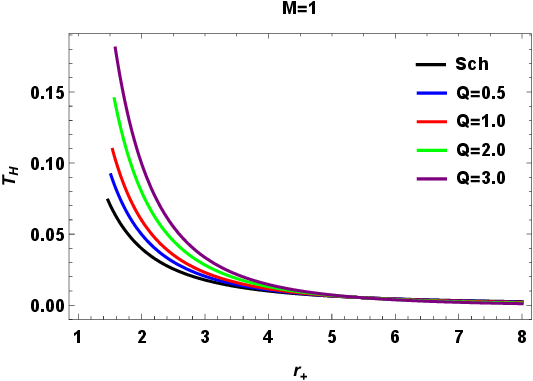}
\caption{Temperature $T_H$ via horizon $r_+$ with varying charge $Q$ and fixed mass $M=1$.}\label{NT1}
\end{figure}

The temperature behaviour for a Horndeski gravity-like hairy BH in the domain $1\leq r_+\leq 8$ and fixed mass $M=1$ is shown in figure \ref{NT1}.
With the rising horizon, the temperature falls exponentially from its peak to an asymptotically flat condition. This state meets Hawking's phenomenon (the BH's small size when prominent radiation is emitted) and shows the physical state of a BH. At non-zero boundaries, low charge values are associated with low temperatures, whereas large charge values are related to high temperatures, demonstrating the precise relationship between charge and temperature.

\section{Tunneling radiation in the Horndeski gravity-like hairy black hole}

The discovery of BH radiation, often known as Hawking radiation, was a significant milestone in establishing a relationship between GR, quantum mechanics, and statistical mechanics. The researchers have studied the nature of BH using quantum mechanics and thermodynamics. The Hawking spectrum has been studied as a quantum tunneling effect of relativistic particles from a BH. The tunneling process of particles from BHs has been widely investigated in different dimensions. String and loop quantum gravity theories show a minimum observable length on the Planck scale, making them ideal choices. A term of such a minimum length results in the GUP. The GUP can be stated \cite{A40} as
\begin{equation}
\Delta y \Delta P \geq \frac{h}{2}\left[1+ \delta(\Delta P)^{2}\right],
\end{equation}
with $\delta =\frac{\delta_{0}}{M_{p}^{2}}$, the Planck mass denoted as $M_{p}$, while the dimensionless parameter is represented by $\delta_{0}$. Then, the modified commutation condition gets
\begin{equation}
\left[y_{\mu}, P_{\nu}\right]=\textrm{i}h\rho_{\mu\nu}[1+\delta P^{2}],
\end{equation}
with the modified position and momentum operators $y_{\mu}$ and $P_{\mu}$, are defined as
\begin{equation}
y_{\mu}=y_{0\mu},~~~P_{\mu}=P_{0\mu}(1+\delta P^{2}_{0\mu}),
\end{equation}
where the commonly used position $y_{0\mu}$ and momentum operators $P_{0\mu}$ often obey the commutation condition $[y_{0\mu}, P_{0\nu}]=\textrm{i}h\rho_{\mu\nu}$. Modified connections have a significant role in the field of physics. Researchers have recently used the GUP to study the thermodynamic aspects of BHs using particle tunneling. The GUP framework \cite{{A31}} modifies the Lagrangian equations to account for quantum gravity impacts. The improved Hawking temperature of different $3+1$ and higher-dimensional BHs was calculated using particle tunneling using these modified relativistic wave equations \cite{A41}. For this reason, we will examine the Hawking radiation of the Horndeski gravity-like hairy BH in four dimensions via the tunneling process of vector particles while considering the impact of the GUP. The semi-classical approach is that both absorbing and radiating particles can do so.

We explore the improved boson tunneling particles at the BH horizon using Lagrangian action with GUP impact. The field expression represents the vector motion of particles (spin-$1$) in spacetime with the charge parameter, mass, and horizon radius related to vector fields taken into account.
Horndeski's theory is based on scalar-tensor gravity, but adding vector fields allows for more complex interactions. The interaction of scalar fields with vector fields can affect spacetime geometry in ways that scalar fields alone cannot. This leads to more generic solutions to the gravitational field equations, such as novel BH solutions and cosmological models.
In the presence of vector fields, we may investigate vector-induced changes to the gravitational field, such as vector hairs near BHs. This can lead to solutions that challenge the standard no-hair theorem of GR and shed light on potential new gravitational phenomena. The Horndeski framework's use of vector fields and GUP effects is noteworthy because it results in a more generic, quantum-corrected explanation of gravity. This broadens the solutions for BHs and cosmological models and introduces significant quantum corrections that alter our knowledge of spacetime on small scales. These modifications broaden Horndeski's theory's scope, allowing for a more in-depth investigation of the behavior of gravity, quantum fields, and the cosmos.The Lagrangian action \cite{A4, A41, A42} with vector field $\Phi_\mu$, improved by GUP stated as
\begin{equation}
\L^{GUP}=-\frac{1}{2}(D^{+}_{\mu}\Phi^{+}_{\nu}
-D^{+}_{\nu}\Phi^{+}_{\mu})
(D^{-\mu}\Phi^{-\nu}-D^{-\nu}\Phi^{-\mu})+
\frac{m^2}{h^{2}}\Phi^{+}_{\mu}\Phi^{-\nu}-\frac{\textrm{i}}{h}F^{\mu\nu}\Phi^{+}_{\mu}\Phi^{-}_{\nu},
\end{equation}
with $F^{\mu\nu}=\nabla^{\mu}V^{\nu}-\nabla^{\nu}V^{\mu}$ and $D^{\pm}_{\mu}=\nabla_{\mu}\pm\frac{\textrm{i}qV_{\mu}}{h}$, $\nabla_{\mu}$, $V_{\mu}$ and $q$  show the covariant derivative, electromagnetic field and particle charge, respectively. To check the spectrum of tunneling pattern for vector motion of particles in a Horndeski gravity-like hairy BH. The expression of the field for the scenario of GUP \cite{A43} is followed as
\begin{eqnarray}
&&\partial_{\mu}(\sqrt{-g}\Phi^{\nu\mu})+
\sqrt{-g}\frac{m^{2}}{h^{2}}\Phi^{\nu}+\sqrt{-g}\frac{\textrm{i}}{h}q V_{\mu}\Phi^{\nu\mu}+
\sqrt{-g}\frac{\textrm{i}}{h}qF^{\nu\mu}\Phi_{\mu}+\delta h^{2}
\partial_{0}\partial_{0}\partial_{0}(\sqrt{-g}g^{00}\Phi^{0\nu})\nonumber\\ &&-
\delta h^{2}
\partial_{i}\partial_{i}\partial_{i}(\sqrt{-g}g^{ii}\Phi^{i\nu})=0,\label{L}
\end{eqnarray}
here $\Phi^{\mu\nu}$, $g$, and $m$ are the anti-symmetric tensor, coefficient matrix, and particle mass, respectively. The $\Phi_{\nu\mu}$ is defined as
\begin{equation}
\Phi_{\nu\mu}=(1-\delta h^{2}\partial^{2}_{\nu})\partial_{\nu}\Phi_{\mu}-
(1-\delta h^{2}\partial^{2}_{\mu})\partial_{\mu}\Phi_{\nu}+
(1-\delta h^{2}\partial^{2}_{\nu})\frac{\textrm{i}}{h}q V_{\nu}\Phi_{\mu}
-(1-\delta h^{2}\partial^{2}_{\mu})\frac{\textrm{i}}{h}q V_{\mu}\Phi_{\nu},\nonumber
\end{equation}
with $\delta$ is the positive parameter of dimensionless (parameter of gravity or parameter of GUP). The tunneling outside and inside of boson particles have the same field action, but their motions are opposite. Next, we will compute particle tunneling for both boson and anti-boson particles. The $\Phi^{\mu}$ elements and $\Phi^{\mu\nu}$ elements are defined as
\begin{eqnarray*}
\Phi^{0}&=&\frac{\Phi_{0}}{g_{tt}},~~~\Phi^{1}=
g_{rr}\Phi_{1},~~~\Phi^{2}=\frac{\Phi_{2}}{r^{2}},~~~
\Phi^{3}=\frac{\Phi_{3}}{r^{2}\sin^{2}\theta},~~~
\Phi^{01}=\frac{g_{rr}\Phi_{01}}{g_{tt}},~~~\Phi^{02}=
\frac{\Phi_{02}}{g_{tt}r^{2}},~~~\\
\Phi^{03}&=&\frac{\Phi_{03}}{g_{tt}r^{2}\sin^{2}\theta},~~~
\Phi^{12}=\frac{g_{rr}\Phi_{12}}{r^{2}},~~
\Phi^{13}=\frac{g_{rr}\Phi_{13}}{r^{2}\sin^{2}\theta},~~
\Phi^{23}=\frac{\Phi_{23}}{r^{4}\sin^{2}\theta}.
\end{eqnarray*}
Ignore the impact of particle self-gravitation and consider a vector particle radiating from Horndeski gravity like hairy BH. The tunneling radiated particles are computed through the Hamilton-Jacobi strategy since we are only concerned with leading throughout the semi-classical strategy. The WKB approximation's formulation is explained as follows:
\begin{equation}
\Phi=d_{\nu}\exp\left[\frac{\textrm{i}}{h}D_{0}(t,r,\theta,\psi)+\sum_{i=1}^{i=n} h^{i}D_{i}(t,r,\theta,\psi)\right].\label{wkb1}
\end{equation}
To get an equation set stated by substituting equation (\ref{wkb1}) into equation (\ref{L}) with $i=1, 2, 3,...$ and discarding the terms of higher order, we get the equation set given as
\begin{eqnarray}
&&g_{rr}[d_{1}(\partial_{0}D_{0})
(\partial_{1}D_{0})+d_{1}\delta(\partial_{0}D_{0})^{3}(\partial_{1}D_{0})
-d_{0}(\partial_{1}D_{0})^{2}-d_{0}(\partial_{1}D_{0})^{4}\delta+q Vd_{1}(\partial_{1}D_{0})
+q Vd_{1}\delta(\partial_{1}D_{0})(\partial_{0}D_{0})^{2}]\nonumber\\
&&+\frac{1}{r^{2}}
[d_{2}(\partial_{0}D_{0})(\partial_{2}D_{0})
+\delta d_{2}(\partial_{0}D_{0})^{3}(\partial_{2}D_{0})
-d_{0}(\partial_{2}D_{0})^{2}-\delta d_{0}(\partial_{2}D_{0})^{4}
+qVd_{2}(\partial_{2}D_{0})+
\delta qVd_{2}(\partial_{0}D_{0})^{2}
(\partial_{2}D_{0})]\nonumber\\
&&+\frac{1}{r^{2}\sin^{2}\theta}
[d_{3}(\partial_{0}D_{0})(\partial_{3}D_{0})
+\delta d_{3}(\partial_{0}D_{0})^{3}(\partial_{3}D_{0})+
d_{0}(\partial_{3}D_{0})^{2}
+\delta d_{0}(\partial_{3}D_{0})^{4}
+qVd_{3}(\partial_{3}D_{0})+\delta d_{3}qV
(\partial_{0}D_{0})^{2}(\partial_{3}D_{0})]\nonumber\\
&&-m^{2}d_{0}=0, \label{31}\\
&&\frac{-1}{g_{tt}}[d_{0}(\partial_{0}D_{0})(\partial_{1}D_{0})
+d_{0}\delta (\partial_{0}D_{0})(\partial_{1}D_{0})^{3}
-d_{1}(\partial_{0}D_{0})^{2}-
d_{1}\delta (\partial_{0}D_{0})^{4}-qVd_{1}(\partial_{0}D_{0})-\delta
qVd_{1}(\partial_{1}D_{0})^{2}(\partial_{0}D_{0})]\nonumber\\
&&+\frac{1}{r^{2}}[d_{2}(\partial_{1}D_{0})(\partial_{2}D_{0})+\delta d_{2}(\partial_{1}D_{0})^{3}(\partial_{2}D_{0})
-d_{1}(\partial_{2}D_{0})^{2}-\delta d_{1}(\partial_{2}D_{0})^{4}]
+\frac{1}{r^{2}\sin^{2}\theta}[d_{3}(\partial_{1}D_{0})
(\partial_{3}D_{0})\nonumber\\
&&+d_{3}\delta
(\partial_{1}D_{0})^{3}
(\partial_{3}D_{0})
-d_{1}(\partial_{3}D_{0})^{2}-d_{1}\delta
(\partial_{3}D_{0})^{4}]
-m^{2}d_{1}-\frac{1}{g_{tt}}qV[d_{0}
(\partial_{1}D_{0})+\delta d_{0}(\partial_{1}D_{0})^{3}-d_{1}(\partial_{0}D_{0})\nonumber\\
&&-\delta d_{1}(\partial_{0}D_{0})^{3}-d_{1}qV
-qV\delta d_{1}
(\partial_{1}D_{0})^{2}]
=0,\label{41}\\
&&\frac{1}{g_{tt}}[d_{0}(\partial_{0}D_{0})(\partial_{2}D_{0})
+\delta d_{0}(\partial_{0}D_{0})(\partial_{2}D_{0})^{3}
-d_{2}(\partial_{0}D_{0})^{2}-\delta d_{2}(\partial_{0}D_{0})^{4}
-qV_{0}(\partial_{0}D_{0})d_{2}
-qV_{0}(\partial_{0}D_{0})^{3}d_{2}\delta
]\nonumber\\
&&-g_{rr}
[d_{2}
(\partial_{1}D_{0})^{2}+\delta d_{2}(\partial_{1}D_{0})^{4}
-d_{1}(\partial_{1}D_{0})(\partial_{2}D_{0})-\delta d_{1}
(\partial_{1}D_{0})(\partial_{2}D_{0})^{3}]+\frac{1}{r^{2}\sin^{2}\theta}
[d_{3}(\partial_{2}D_{0})(\partial_{3}D_{0})\nonumber\\
&&+\delta d_{3}(\partial_{2}D_{0})^{3}(\partial_{3}D_{0})
-d_{2}(\partial_{3}D_{0})^{2}-
\delta (\partial_{3}D_{0})^{4}d_{2}]
-\frac{Vq}{g_{tt}}[(\partial_{2}D_{0})d_{0}
+\delta (\partial_{2}D_{0})^{3}d_{0}-
(\partial_{0}D_{0})d_{2}-\delta (\partial_{0}D_{0})^{3}d_{2}\nonumber\\
&&+qVd_{2}+\delta d_{2}qV(\partial_{0}D_{0})^{2}]
-d_{2}m^{2}=0, \label{51}\\
&&+\frac{1}{g_{tt}}[(\partial_{0}D_{0})(\partial_{3}D_{0})d_{0}
+\delta (\partial_{0}D_{0})(\partial_{3}D_{0})^{3}d_{0}
-(\partial_{0}D_{0})^{2}d_{3}-\delta (\partial_{0}D_{0})^{4}d_{3}
-qV(\partial_{0}D_{0})d_{3}
-qV(\partial_{3}D_{0})^{2}(\partial_{0}
D_{0})d_{3}\delta]\nonumber\\
&&+g_{rr}[d_{3}
(\partial_{1}D_{0})^{2}+\delta d_{3}(\partial_{1}D_{0})^{4}
-d_{1}(\partial_{3}D_{0})(\partial_{1}D_{0})-\delta d_{1}
(\partial_{1}D_{0})(\partial_{3}D_{0})^{3}]+\frac{1}{r^{2}}
[d_{3}(\partial_{2}D_{0})^{2}+\delta d_{3}(\partial_{2}D_{0})^{4}\nonumber\\
&&-d_{2}(\partial_{2}D_{0})
(\partial_{3}D_{0})-
\delta d_{2}(\partial_{3}D_{0})^{3}(\partial_{2}D_{0})]
+\frac{qV}{g_{tt}}[d_{0}(\partial_{3}D_{0})
+\delta d_{0}(\partial_{3}D_{0})^{3}-d_{3}
(\partial_{0}D_{0})
-\delta d_{3}(\partial_{0}D_{0})^{3}
-d_{3}qV\nonumber\\
&&-\delta d_{3}qV(\partial_{3}D_{0})^{2}]
-d_{3}m^{2}=0.\label{61}
\end{eqnarray}
To separate the variables, use the given approach in the form as follows
\begin{equation}
D_{0}=-\Omega t+I(r)+R\phi+S(\theta),\label{RRA}
\end{equation}
with $\Omega=(E-Jv)$, the angular velocity, energy, and angular momentum are represented by the variables $v$, $E$, and $J$, respectively. In this case, two arbitrary functions are $I(r)$ and $S(\theta)$. By inserting equations, one can obtain the matrix equation. In the expressions ~(\ref{31})--(\ref{61}) into Eq. (\ref{RRA}) as
\begin{equation*}
H(d_{0},d_{1},d_{2},d_{3})^{T}=0,
\end{equation*}
whose entries are computed as follows and this yields a $'4 \times 4'$ order matrix indicated as "$H$". The components of "$H$" can be expressed as
\begin{eqnarray}
H_{00}&=&g_{rr}(\dot{I}^{2}+\delta\dot{I}^{4})-\frac{R^{2}+\delta R^{4}}{r^{2}}
+\frac{\dot{S}^{3}+\delta\dot{S}^{4}}{r^{2}\sin^{2}\theta}-m^{2},~~~~~
H_{01}=-g_{rr}(\dot{I}\Omega+\delta\dot{I}\Omega^{3})
+g_{rr}(\dot{I}qV+\delta\dot{I}qV\Omega^{2}),\nonumber\\
H_{02}&=&-\frac{\Omega R+\delta \Omega R}{r^{2}}+
\frac{qVR+\delta \Omega^{2}qVR}{r^{2}},~~~~~
H_{03}=-\frac{\dot{S}\Omega+\delta\dot{S}\Omega^{3}}
{r^{2}\sin^{2}\theta}+\frac{qV\dot{S}+\delta qV\dot{S}\Omega^{2}}{r^{2}\sin^{2}\theta},\nonumber\\
H_{10}&=&\frac{\Omega\dot{I}+\delta \Omega\dot{I}^{3}}
{g_{tt}}-\frac{qV\dot{I}+\delta qV\dot{I}^{3}}{g_{tt}},~~~~~H_{12}=\frac{\dot{I}R+\delta\dot{I}^{3}R}{r^{2}},~~~~~
H_{13}=\frac{\dot{S}\dot{I}+\dot{S}\delta\dot{I}^{3}}{r^{2}\sin^{2}\theta},\nonumber\\
H_{11}&=&\frac{\Omega^{2}+\delta \Omega^{4}}{g_{tt}}+
\frac{\Omega qV-\delta\dot{I}
\Omega qV}{g_{tt}}
-\frac{{R}^{2}-\delta {R}^{4}}{r^{2}}-\frac{\dot{S}^{2}-\delta\dot{S}^{4}}{r^{2}\sin^{2}\theta}
-m^{2}-\frac{1}{g_{tt}}qV[\Omega+\delta \Omega^{3}
-qV-\delta qV\dot{I}^{2}],\nonumber\\
H_{20}&=&-\frac{R\Omega+\delta R^{3}\Omega}{g_{tt}}
-qV\frac{R+\delta R^{3}}{g_{tt}},~~
H_{22}=-\frac{1}{g_{tt}}[-\Omega^{2}-\delta \Omega^{4}+qV
\Omega+qV\delta \Omega^{3}]
-(\dot{I}^{2}+\delta\dot{I}^{4})g_{rr}
-\frac{\dot{S}^{2}+\delta\dot{S}^{4}}{r^{2}\sin^{2}\theta}-m^{2},\nonumber\\
H_{21}&=&g_{rr}(\dot{I}R+\delta\dot{I}R^{3}),~~~~~
H_{23}=\frac{R\dot{S}+\delta R^{3}\dot{S}}{r^{2}\sin^{2}\theta},~~~
H_{30}=\frac{-1}{g_{tt}}[\Omega\dot{S}+\delta \Omega\dot{S}^{3}]
+\frac{qV\dot{S}+qV\delta\dot{S}^{3}}{g_{tt}},\nonumber\\
H_{31}&=&(-\dot{I}\dot{S}-\delta\dot{I}\dot{S}^{3})g_{rr},~~~~~
H_{32}=\frac{-R\dot{S}-\delta R\dot{S}^{3}}{r^{2}},\nonumber\\
H_{33}&=&-\frac{1}{g_{rr}}[\Omega^{2}+\delta \Omega^{4}-
\Omega qV-\delta \Omega qV\dot{S}^{3}]+
(\dot{I}^{2}+\delta\dot{I}^{4})g_{rr}-\frac{R^{2}+\delta R^{4}}{r^{2}}
+\frac{qV}{g_{tt}}[\Omega+\delta \Omega^{3}
-qV-\delta qV\dot{S}^{3}]-m^{2},\nonumber
\end{eqnarray}
where $\Omega=\partial_{t}D_{0}$, $\dot{I}=\partial_{r}D_{0}$, $\dot{S}=\partial_{\theta}D_{0}$ and $R=\partial_{\phi}D_{0}$. We consider $\mid\textbf{H}\mid=0$ when solving these equations for a non-trivial answer, and the corresponding result  imaginary of $I$ appears as
\begin{equation}\label{w1}
I^{\pm}=\pm \int\sqrt{\frac{(\Omega-qV)^{2}+Z_{1}(1+\frac{Z_{2}}{Z_{1}}\delta)}{g_{rr}}}dr,
\end{equation}
where $+$ and $-$ represent the emitting and absorbing particles, respectively. The functions $Z_1$ and $Z_2$ are described as
\begin{equation}
Z_{1}=\frac{R^{2}}{r^{2}},~~~~~
Z_{2}=\frac{\delta \Omega^{4}}{g_{tt}}-\frac{\delta \Omega qV\dot{S^{3}}}{g_{tt}}-
(\delta\dot{I}^{4})g_{rr}+\delta\frac{R^{4}}{r^{2}}
-\frac{qV}{g_{tt}}[\delta \Omega^{3}-\delta qV\dot{S}^{3}] +m^{2}\nonumber,
\end{equation}
This shows the angular velocity at the BH's horizon. The contour should be an infinite semi-circle below the pole $r=r_{+}$ in the normal coordinate framework to radiate particles from the horizon. In a similar way, the absorbing of particle contour appears over the pole. The field theory is defined as multiplying it by the complex conjugate for the semi-classical emission of tunneling. The portion of the pathway that starts BH outside and proceeds to the observation can be emitted since it will be real or imaginary and will not be viewed when calculating the final emission of tunneling. Thus, the contour around the field is the sole part of the field trajectory equation, which implies rise to the emission of tunneling. An integration of the solution (\ref{w1}) over the pole gets
\begin{equation}
I^{\pm}
=\pm \textrm{i}\pi\frac{(\Omega-qV)}{2K(r_{+})}[1+X\delta].
\end{equation}
The surface gravity is denoted by $K(r_{+})$. It is important to note that $X=6\left(\frac{J^{2}_{\psi}\csc^{2}\theta+J^{2}_{\theta}}{r^{2}_{+}}+m^{2}\right)>0$ indicates the kinetic energy component along the horizon surface at the moment of emission position. We examine the numerically corresponding general idea inferred that the interior contour is situated in the plane of the lower half and undergoes multiplication by the negative symbol, as in $I^{+}(r_{+})=-I^{-}(r_{-})$. The quantum physical technique known as Hawking emission is explained in the following way: First, particles and anti-particles, which are close to the horizon of energy particles, have positively charged particles that tunnel outside the horizon and are seen by their external observer as Hawking emissions, while the opposite particle falls within the horizon. Secondly, the particle pair forms within the horizon. The energy of their counterpart of negative continues to exist behind; the positive energy particles can cross the energy barrier by tunnel to infinity. A loses BH mass when it exits energy in the sense of Hawking energy, which includes a quantum tunneling mechanism in which particles have a finite probability of radiating via the BH horizon and classical particles cannot do. In this case, we examine the emission of tunneling based on the imaginary component of the particle action emitted by the horizon. We study the emission of tunneling for pathways of classical forbidden fields from inside horizon to outside horizon. Utilizing the WKB approximation, the expression $\Gamma\approx[-2I]~(\textit{h=1})$ can be obtained according to the classical framework for vector particle action tunneling within the BH horizon as pathways that exceed leading order. The $\Gamma(I^{+})$ is explained \cite{A44} by
\begin{equation}
\Gamma(I^{+})=\frac{\Gamma_{\textrm{emission}}}{\Gamma_{\textrm{absorption}}}=\exp\left[-\frac{2\pi(\Omega-qV)}{K(r_{+})}\right][1+X\delta],\label{TUN}
\end{equation}
with 
\begin{equation}\label{K}
K(r_{+})=\frac{g'_{tt}(r)}{2}=\frac{M}{ r_{+}^{2}}+\frac{Q}{2r_{+}^{2}}-\frac{Q}{ 2r_{+}^{2}}\ln\left(\frac{r_{+}}{2M}\right).
\end{equation}
In the absence of this charge field, it reduces to the Schwarzschild of the vacuum spacetime, and it would prove interesting to explore the tunneling condition from such a BH. Several BH solutions incorporate the charge variable and look at its physical impact in the space of colliding waves. Therefore, we examined the physical process in the context of potential, charge, energy and angular momentum, i.e., $2I=\beta(\Omega-qV)+O(\Omega-qV)$, to find the improved Hawking temperature ($T'_{H}$). It indicates that, in the tunneling technique, the emission rate takes the Boltzmann factor $T(I^{+})\approx \textrm{e}^{-(\Omega-qV)\beta}$, where $\frac{1}{T'_{H}}=\beta$, up to first order. Concepts of higher order represent the self-interaction impacts based on momentum-energy conservation. The corresponding $T'_{H}$ on the horizon is given by
\begin{equation}\label{TH2}
T'_{H}=\left[\frac{M}{2\pi r_{+}^{2}}+\frac{Q}{4\pi r_{+}^{2}}-\frac{Q}{4\pi r_{+}^{2}}\ln\left(\frac{r_{+}}{2M}\right)\right]
(1-X\delta).
\end{equation}

The improved temperature depends on the BH mass $M$, charge parameter $Q$, GUP correction parameter $\delta$, and radial coordinate of the horizon $r_+$. It's also crucial to note that the $T'_{H}$ becomes the $T_{H}$ (\ref{TH1}) of Horndeski gravity-like hairy BH when the GUP modification effects (i.e., $\delta=0$) are neglected.
If we disregard the quantum corrections, tunneling depends on these emitting particles and space-time properties. The tunneling probabilities are used to compute the temperature. With the quantum gravity effects described by the GUP, a BH Hawking temperature typically decreases. The GUP introduces modifications to the near-horizon geometry, which affects the quantum tunneling process that produces Hawking radiation, and the GUP plays an important influence on thermodynamics in \cite{A45, A46}.

By implying that BHs do not entirely destroy information due to quantum gravitational effects, current theoretical research suggests that the GUP may provide some degree of information recovery through Hawking radiation, which would be corrected by the GUP corrections. In other words, the GUP may introduce a mechanism to encode more information within the emitted radiation, thereby limiting the information loss issue.

\begin{figure}[H]
\centering
\includegraphics[width=6cm,height=6cm]{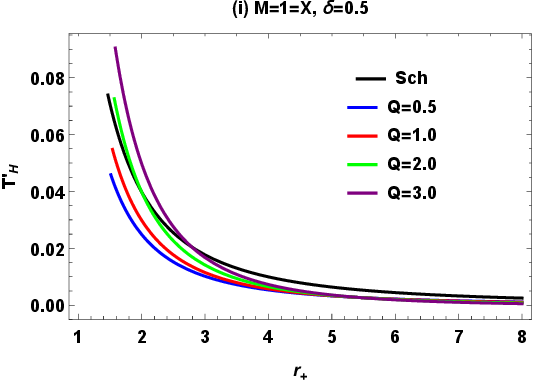}\includegraphics[width=6cm,height=6cm]{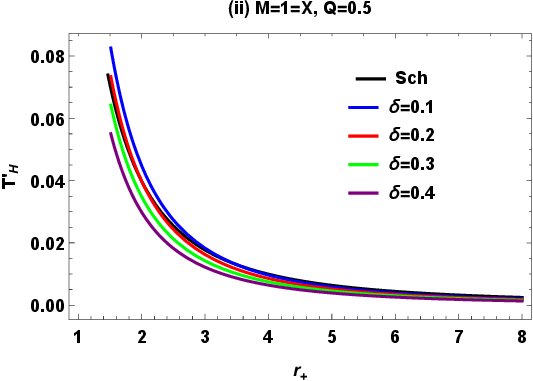}
\caption{Corrected temperature $T'_H$ versus horizon 
radius $r_+$ for constant values of BH mass $M=1$ and arbitrary parameter $X=1$ with variation of charge $Q$ and gravity parameter $\delta$.}\label{T1}
\end{figure}

Figure \ref{T1} gives the graphical conduct of corrected temperature for Horndeski gravity-like hairy BH for constant $M=1=X$ in the region $1\leq r_+\leq 8$.

The behavior for changes in charge $Q$ and fixed gravity parameter values $\delta=0.5$ is depicted in the left plot. It is evident that the charge parameter affects temperature directly; at non-zero horizons, we find low temperatures for small charge parameter values and high temperatures for big ones. We can observe that the temperature is exponentially decreasing from a maximum point with rising $r_+$, this condition satisfies the Hawking phenomenon, which states that the BH temperature decreases as the horizon radius increases or we can say that with the emission of large number of radiations, the size of the BH reduces. So, we can observe a maximum temperature at a small value of horizon radius.

The right plot shows the behavior for different gravity parameters $\delta$ and a fixed charge value $Q=0.5$.
The temperature lowers exponentially as the horizon rises. This state reflects the physical state of a BH and satisfies Hawking's phenomena. The temperature is low when the gravity parameter is high and high when the gravity parameter is small. When BH's corrected temperature drops to zero ($T'_H=0$), it means that the BH radiates no net energy. The BH achieves thermodynamic stability, reaching a point where quantum and classical effects cancel each other out. It frequently indicates equilibrium, a possible phase transition, or a critical or extreme state without further radiation. This condition is essential for investigating complex quantum gravity and BH thermodynamics ideas.

It is important to note here that the quantum corrections slow down the growth in Hawking temperature. We may visually analyze this fact and see that the temperature in the case of quantum corrections is lower than the standard temperature.

\section{Entropy Corrections for Horndeski gravity like hairy BLACK HOLE}
This section discusses the \textit{Bekenstein-Hawking entropy} and its relation to the quantum loop expansion theory correction factor. The logarithmic term occurs as an additional loop in conventional of BH's entropy. According to \cite{V1}, BH entropy plays a crucial role in primordial BHs. To give the small BHs enough time to evaporate in their current state completely, it establishes a primordial BH wit minimum mass.
We examine entropy corrections for Horndeski's gravity-like hairy BH. Banerjee and Majhi \cite{V2}-\cite{V4} used the null geodesic technique to explore entropy adjustments. Using the \textit{Bekenstein-Hawking entropy} formula for order of first corrections, we can calculate the corrected entropy of Horndeski gravity-like hairy BH \cite{V5}.
We use the formulas of ${T'}_H$ and standard entropy $\mathbb{S}_{o}$ to calculate the logarithmic corrected entropy for Horndeski gravity-like hairy BH using the formula as given below \cite{d1}
\begin{equation}
\mathbb{S}_{H}=\mathbb{S}_{o}-\frac{1}{2}\ln\Big|{T'}^2_H \mathbb{S}_{o}\Big|+...~.\label{vv}
\end{equation}
To analyses the corrected entropy for Horndeski gravity like-hairy BH, firstly, we evaluate the standard entropy for Horndeski gravity like-hairy BH using the given formula
\begin{equation}
\mathbb{S}_{o}=\frac{A_H}{4},\label{v2}
\end{equation}
where $A_H$ represents the standard area for Horndeski gravity like-hairy BH and can be calculated as follows
\begin{equation}
A_H=\int_{0}^{2\pi}\int_{0}^{\pi}\sqrt{g_{\theta\theta}g_{\phi\phi}}\textrm{d}\theta \textrm{d}\phi=4\pi r^2.\label{v1}
\end{equation}
To calculate the standard entropy for Horndeski gravity-like-hairy BH, use the values from Eq. (\ref{v1}) into Eq. (\ref{v2}) and obtain
\begin{equation}
\mathbb{S}_{o}=\pi r^2.\label{n1}
\end{equation}
By using the values from Eqs. (\ref{TH2}) and (\ref{n1}) into Eq. (\ref{vv}), we get the corrected entropy for Horndeski gravity-like hairy BH in the provided form as
\begin{eqnarray}
\mathbb{S}_{H}&=&\pi r_{+}^2-\frac{1}{2}\ln\left|\frac{1}{\pi r_{+}^2}\left[\left(\frac{M}{2}+\frac{Q}{4}-\frac{Q}{4}\ln\left(\frac{r_+}{2M}\right)\right)
\left(1-X\delta\right)\right]^2\right|+...,\label{b2}
\end{eqnarray}

The Eq. (\ref{b2}) gives the corrected entropy for Horndeski gravity like-hairy BH. The corrected entropy depends on BH mass $M$, charge $Q$, horizon radius $r_+$, kinetic energy parameter $X$, and gravity parameter $\delta$.

\begin{figure}[H]
\centering
\includegraphics[width=6cm,height=6cm]{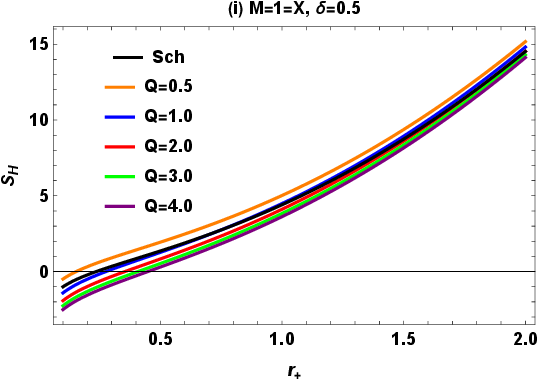}\includegraphics[width=6cm,height=6cm]{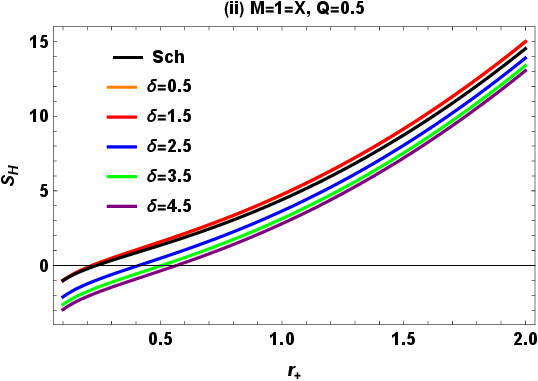}
\caption{Modified entropy $S_H$ w.r.t BH horizon $r_+$ for changes of charge parameter $Q$ and gravity parameter $\delta$.}\label{E1}
\end{figure}

Figure \ref{E1} represents the behavior of corrected entropy for Horndeski gravity like-hairy BH for constant mass and arbitrary parameter $M=1=X$.

The left panel demonstrates the behavior for variations of charge $Q$ and constant values of gravity parameter $\delta=0.5$. For every charge value $Q$, the entropy rises as $r_+$ increases. The trajectory of this increase is steady and consistent. The curves are tightly spaced for lower values of $r_+$, suggesting that $S_H$ is less affected by the BH charge $Q$ in this area. Higher $Q$ values (such as $Q=4.0$) produce less entropy than lower $Q$ values (such as $Q=0.5$) as $r_+$ rises. With $Q=0.5$ retaining the greatest $r_+$, the entropy differences across curves become apparent when $r_+>1$. The rise of entropy appears to be suppressed by a greater charge $Q$, indicating a bigger charge field effect on the thermodynamic characteristics and horizon structure. The black curve shows the Schwarzschild solution $(Q = 0)$. It continuously stands above the $Q>0$ curves for $r_+>0.5$, emphasizing charge $Q$ entropy-reducing impact.

The right panel shows the behavior for changes of gravity parameter $\delta$ and fixed values of charge $Q=0.5$. As $r_+$ rises, $S_H$ increases, and $\delta$ affects the growth rate. The entropy $S_H$ is not greatly affected by $\delta$ for tiny $r_+$; the curves begin closely together close to the origin.
Higher $\delta$ values (e.g., $\delta=4.5$) result in more entropy for bigger $r_+$. In contrast to $\delta=4.5$, the curve for $\delta=1.5$ increases more slowly. More significant $\delta$ significantly increases entropy for bigger BHs, as seen by the widening gap between $S_H$ curves for $r_+ >1$. The $\delta$ parameter significantly impacts BH entropy and may be related to deformation or external scalar fields. Entropy is increased with larger $\delta$, most likely because of changes in spacetime curvature close to the horizon. 
The black curve depicts the entropy of the Schwarzschild black hole, a specific instance with no extra modifications $(\delta=0)$. It exhibits a conventional positive monotonic rise for $r_+$. The parameter $\delta$ appears to regulate a divergence from the conventional Schwarzschild scenario. This might reflect a variation in the BH metric or an additional parameter in an extended gravity theory. A greater $\delta$ decreases entropy for decreasing $r_+$, indicating an influence on the BHs thermodynamic characteristics, potentially due to changes in its microstate structure or quantum corrections. The drop in entropy with rising $\delta$ implies that modified BHs may have fewer microstates than the Schwarzschild scenario. This decrease might indicate a more constrained phase space.

The regions in which corrected entropy is less than zero are in contrast to the conventional second law of thermodynamics, which states that entropy cannot drop in a closed system; negative entropy suggests that the system has less than zero information.
This would indicate that the system had attained a "more order" state than the lowest entropy state, which in classical thermodynamics has no physical significance.

From Fig. \ref{E1}, we conclude that the corrected entropy $S_H$ is amplified by $\delta$, whereas $Q$ tends to reduce it. This suggests that the effects of deformation parameters and electric charge on BH thermodynamics are opposite.
Particularly for greater $r_+$, the effect of $\delta$ on entropy seems more noticeable than that of $Q$.
For tiny horizons, $Q$ and $\delta$ have negligible impacts on entropy, highlighting the dominance of $r_+$ in this regime. For the same range of $r_+$, the right panel indicates a slower rise of $S_H$ than the left panel, demonstrating that $\delta$ significantly increases entropy over $Q$.

\section{Conclusions}
A particle with electro-positive energy in tunneling crosses the horizon and generates an electromagnetic field. According to this scenario, an electro-negative particle in nature burrows into the weave, is absorbed by the BH, and eventually sees its mass decrease and evaporate. As a result, the particles may move in an incoming or outgoing configuration, causing their actions to be real and complex, respectively. The imaginary component of the particle's action, connected to the Boltzmann factor depending on the Hawking temperature, corresponds to the emission rate of tunneling particles from the BH.

In this research, we have obtained the corresponding temperature at which particles tunnel via horizons, expanding the work of charged boson particle tunneling for more generalized Horndeski gravity-like hairy BH in $4$D space. To do this, we have examined the tunneling of boson particles in four dimensions using the field equation and an electromagnetic background. The set of field equations is obtained by applying the WKB approximation to the Proca problem, which we then solved by separating variables. We compute the radial component using the determinant of the coefficient matrix, which is zero. Using surface gravity, We calculated the tunneling probability and the Hawking temperature for BH at the outer horizon; all these values depend on the BHs' defining parameters and quantum gravity. It is important to note that the derived Hawking temperature is only a leading term and does not require calculating the proper solution of the semi-classical technique for the spacetime of the background BH in equilibrium with its Hawking radiation. 

Our analysis leads us to conclude that the Hawking temperature at which particles tunnel through the horizon is independent of the species of particles; in particular, it is independent of the nature of the background BH geometries. The semi-classical effects will show that their tunneling probabilities are the same for particles with different spins (either spin up or spin down) or zero spins. Therefore, the corresponding Hawking temperatures must be the same for all particles. In this scenario, our results are consistent with the claim that the temperature of tunneling particles is independent of the species of the particles, and this result is also valid for different coordinate frames by employing particular coordinate transformations. Therefore, the conclusion remains valid even if background BH geometries are more general.

Moreover, we have analyzed that the improved temperature depends on the BH mass, charge parameter, GUP correction parameter, and radial coordinate of the horizon. It's also crucial to note that the improved temperature becomes the standard temperature (\ref{TH1}) of Horndeski gravity-like hairy BH when the GUP modification effects (i.e., $\delta=0$) are disregarded.
If we disregard the quantum corrections, tunneling becomes dependent on these emitting particles and space-time properties. Furthermore, the Schwarzschild BH of temperature 
\Big(i.e., $T_{H}=\frac{1}{8\pi M}$\Big) is recovered when $r_{+}=2M$ and the charge is disregarded.

From graphical analysis, we concluded that the charge parameter affects temperature directly; at non-zero horizons, we observed low temperatures for small charge parameter values and high temperatures for big ones. We observed that the temperature exponentially decreases from a maximum point with rising $r_+$, this condition satisfies the Hawking phenomenon, which states that the BH temperature decreases as the horizon radius increases or we can say that with the emission of large number of radiations, the size of the BH reduces. So, we observed a maximum temperature at a small value of horizon radius. Moreover, The temperature is low when the gravity parameter is high, and the temperature is high when the gravity parameter is small. It is also worth noting that $T'_H=0$ is the point at which correction terms cancel out of the original temperature term, making it a critical point. We also observed from our graphical analysis that in the presence of quantum gravity, the corrected temperature is lower than the standard one (in the absence of the quantum gravity parameter). Therefore, we can conclude that the quantum corrections ceased the rise in temperature.

In the end, we investigated the corrected entropy for Horndeski gravity-like hairy BH using corrected temperature and standard entropy. We also studied the graphical effects of charge and gravity parameters on corrected entropy.
Entropy increases exponentially as the horizon rises. For large BH, the entropy grows with increasing horizons, indicating the BH stable form. We also observed that charge has an inverse impact on entropy. As we increase the values of charge, the entropy decreases. The corrected entropy drops as the gravity parameter increases, implying that the gravity parameter influences corrected entropy in the other direction.

\section*{Acknowledgement}

The paper was funded by the National Natural Science Foundation of China 11975145.

\end{document}